\documentclass[onecolumn]{IEEEtran}

\ifCLASSINFOpdf

\else

\fi
\usepackage[T1]{fontenc}
\usepackage{setspace}
\usepackage{subcaption} 
\usepackage{epstopdf}
\usepackage[official]{eurosym}
\usepackage{url}
\usepackage[cmex10]{amsmath}
\usepackage{cases}
\usepackage{color}
\usepackage{xcolor}
\usepackage{units}
\usepackage{epstopdf}
\usepackage{graphicx}
\usepackage{tabularx}
\usepackage{booktabs}
\usepackage{multirow}
\usepackage{algorithm}
\usepackage{algpseudocode}

\usepackage{array}
\newcolumntype{P}[1]{>{\centering\arraybackslash}p{#1}}
\newcolumntype{M}[1]{>{\centering\arraybackslash}m{#1}}
\usepackage{amsmath}
\usepackage[cmintegrals]{newtxmath}
\usepackage{diagbox}
\usepackage{makecell}
\usepackage{algorithm}
\usepackage{algpseudocode}
\usepackage{algorithmicx}
\usepackage{tikz}
\usetikzlibrary {positioning}
\usetikzlibrary{automata,arrows,positioning,calc}
\definecolor {processblue}{cmyk}{0.96,0,0,0}
\usepackage{filecontents,lipsum}
\usepackage[noadjust]{cite}
\usepackage[acronym]{glossaries}
\usepackage[final]{changes}
\makeatletter
\AddToHook{cmd/added/before}{\def\Changes@AuthorColor{blue}}
\AddToHook{cmd/deleted/before}{\def\Changes@AuthorColor{green}}
\AddToHook{cmd/replaced/before}{\def\Changes@AuthorColor{red}}
\makeatother

\newacronym{ETSI}{ETSI}{European Telecommunications Standards Institute}
\newacronym{ILP}{ILP}{Integer Linear Program}
\newacronym{MILP}{MILP}{Mixed Integer Linear Program}
\newacronym{NFV}{NFV}{Network Function Virtualization}

\newacronym{PSN}{PSN}{Physical Network Substrate}

\newacronym{QoS}{QoS}{Quality of Service}

\newacronym{SFC}{SFC}{Service Function Chain}

\newacronym{VNE}{VNE}{Virtual Network Embedding}
\newacronym{VNF}{VNF}{Virtual Network Function}
\newacronym{VNF-FG}{VNF-FG}{Virtual Network Function Forwarding Graph}
\newacronym{VNF-FGE}{VNF-FGE}{Virtual Network Function Forwarding Graph Embedding}

\usepackage{color}

\usepackage{orcidlink}
\newcommand{\cF}{{\cal F}}
\newcommand{\cR}{{\cal R}}
\newcommand{\cS}{{\cal S}}

\newcommand{\cD}{{\cal D}}

\newcommand{\cG}{{\cal G}}
\newcommand{\cV}{{\cal V}}
\newcommand{\cA}{{\cal A}}

\newcommand{\cE}{{\cal E}}

\usepackage{float}
\usepackage{caption}
\usepackage{subcaption}
\usepackage{verbatim}
\usepackage{mathrsfs}
\usepackage{mathtools}

\usepackage{pgfplots}
\pgfplotsset{compat=1.18} 
\usepackage{subcaption}
\usepackage{balance}

\begin{document}

\title{Online Network Slice Deployment across Multiple Domains under Trust Constraints}

\author{Julien Ali~El~Amine\,\orcidlink{0000-0003-1001-189X}, \textit{Senior Member, IEEE}, Nour El Houda Nouar and Olivier Brun
	\thanks{J.~El~Amine is with College of Engineering and Technology, American University of the Middle East, Egaila 54200, Kuwait - Email: julien.el-amine@aum.edu.kw}
    \thanks{N.~Nouwar is with LAAS-CNRS, 7 Av. du Colonel Roche, 31400, Toulouse France - Email: nouar@laas.fr}
\thanks{O.~Brun is with LAAS-CNRS, 7 Av. du Colonel Roche, 31400, Toulouse France - Email: brun@laas.fr}}

\maketitle
\vspace{0.5em}
\begin{center}
\footnotesize
This paper has been accepted for publication in \textit{IEEE Transactions on Network and Service Management}.\\
© IEEE. Personal use of this material is permitted. The final version will be available at IEEE Xplore.
\end{center}
\vspace{0.5em}

\begin{abstract}
	Network slicing across multiple administrative domains raises two coupled challenges: enforcing slice-specific trust constraints while enabling fast online admission and placement decisions. This paper considers a multi-domain infrastructure where each slice request specifies a VNF chain, resource demands, and a set of (un)trusted operators, and formulates the problem as a Node–Link (NL) integer program to obtain an optimal benchmark, before proposing a Path–Link (PL) formulation that pre-generates trust and order-compliant candidate paths to enable real-time operation.
To mitigate congestion, resource prices are made dynamic using a Kleinrock congestion function, which inflates marginal costs as utilization approaches capacity, steering traffic away from hotspots.
Extensive simulations across different congestion levels and slice types show that: (i) PL closely tracks NL with negligible gaps at low load and moderate gaps otherwise, (ii) dynamic pricing significantly reduces blocking under scarce resources, and (iii) PL reduces computation time by about 3×–6× compared to NL, remaining within a few seconds even at high load. These results demonstrate that the proposed PL and dynamic pricing framework achieves near-optimal performance with practical runtime for online multi-domain slicing under trust constraints.
\end{abstract}


\section{Introduction}
\label{Introduction}
Lately, \gls{NFV} has become a compelling approach for delivering flexible, low-cost network services by relocating functions (e.g. firewalls, deep packet inspection, and web proxies) from dedicated hardware appliances to software instances running on standard off-the-shelf servers~\cite{mijumbi2015network}, known as \glspl{VNF}, thereby enabling the on-demand deployment of network slices as isolated logical networks coexisting on the same \gls{PSN} and tailored to specific application requirements.
To realize a network slice, \glspl{VNF} must be strategically deployed on \gls{NFV}-capable network nodes and connected in a particular order to form a \gls{VNF-FG}, which \gls{ETSI} defines as a directed graph of an ordered set of \glspl{VNF} composing an end-to-end service\footnote{Throughout the paper, we use the terms service and slice interchangeably.}~\cite{etsi2013network}.

The logical service graph described by a \gls{VNF-FG} is decoupled from the physical network and must be mapped onto the available \gls{PSN} resources \cite{Herrera2016}. This mapping process must not only fulfill the slice’s Quality of Service (QoS) requirements but also respect the constraints imposed by the underlying infrastructure. Basically, two types of decisions have to be made: (a) where to run the \glspl{VNF}, and (b) how to interconnect them in the physical network, taking into account specific \gls{VNF} ordering requirements. These decisions have to be made for optimal performance and better use of network resources while satisfying a number of technical constraints.

The above problem, referred to as the \gls{VNF-FGE} problem, extends the classical \gls{VNE} problem, which is known to be NP-hard~\cite{lim2023reinforcement}. The \gls{VNF-FGE} problem has been studied over the past years in two main contexts: the offline context, where multiple \glspl{VNF-FG}—known in advance—must be simultaneously deployed in the physical network~\cite{tastevin2017graph, zhong2019cost, guo2020cost, el2022game}, and the online context, where slice requests arrive sequentially without knowledge of future demands, requiring immediate deployment upon arrival~\cite{quang2018single, el2021shortening, taghavian2023fair, esfandyari2025online}.

Nevertheless, existing works on the \gls{VNF-FGE} problem mainly focus on single administrative domains, while the challenges of multi-domain network slice orchestration remain largely overlooked. In a multi-domain context, a single service may span several domains managed by different operators, requiring inter-operator collaboration and information disclosure, while raising confidentiality and integrity concerns as traffic may traverse networks operated by untrusted authorities. For critical services, such collaboration is therefore likely to be restricted to operators with mutual trust.

In this paper, we study the online \gls{VNF-FGE} problem under trust constraints, aiming to reduce the cost and improve resource utilization.
In particular, we consider a multi-domain physical infrastructure with known transmission and capacity resources to accommodate slice requests, where each domain comprises multiple operators.
Moreover, each slice request imposes specific trust constraints on the operators that can handle its traffic. Depending on these constraints, the traffic associated with a given slice may be restricted from traversing certain operators, ensuring that only trusted operators participate in the provisioning of the slice.
Under this model, we focus this paper on studying the online placement of slices with diverse trust and technical requirements.
\par
The main contributions of this paper are as follows:
\begin{itemize}
    \item We formalize online multi-domain slice deployment with explicit trust constraints as a Node–Link (NL) ILP, providing an optimal per-request benchmark.
    \item We propose a Path–Link (PL) formulation that pre-generates trust-compliant candidate paths and solves a reduced ILP, achieving near-optimal blocking performance while cutting computation time by 3–6$\times$.
    \item We introduce a dynamic resource-pricing scheme to steer requests away from highly utilized resources, significantly lowering blocking under scarce-capacity scenarios. 
    \item Extensive simulations over different congestion regimes and slice types validate the approach.
\end{itemize}

The paper is organized as follows. Section~\ref{Related work} reviews related work and highlights similarities and differences with the present study. The network model is described in Section~\ref{Network Model}, followed by the NL MILP formulation for optimal slice placement in Section~\ref{Node-Link Formulation} and the PL formulation in Section~\ref{Path-Link Formulation}. Section~\ref{Dynamic pricing} presents the dynamic resource-pricing scheme, Section~\ref{Simulation and results} reports the numerical results, and Section~\ref{Conclusion} concludes the paper and outlines future research directions.

\section{Related work}
\label{Related work}

The virtual network embedding problem has been extensively studied in the literature; see, e.g. \cite{Herrera2016,zhong2019cost,guo2020cost,Luizelli2018,Pei2019,Li2021,el2022game,assis2025two}. Nevertheless, the research on this problem was for long mainly focused on the case of a single administrative domain with a thorough knowledge of the substrate network. Moreover, most works were focused on the offline setting. In contrast, in the present paper we consider the online version of the problem assuming that the \gls{PSN} is comprised of multiple networks operated by independent administrative entities.

Setting up network services on top of multi-administrative multi-domain networks is much more challenging for several reasons~\cite{Katsalis2016,Rosa2015,Baranda2020,Valcarenghi2018}. First, the orchestration of multi-domain services requires information sharing between different network operators. As they are usually reluctant to disclose detailed information about their network topology and resource availability to third parties, network operators usually only disclose an abstraction of their infrastructure, with a mesh of links between edge nodes (peering nodes). This legacy approach has significant implications in terms of resource discovery and allocation, and several studies have advocated information disclosure policies that enable more effective collaborations between domains~\added{\cite{pentelas2020network,minardi2022virtual,zhang2023multi}}. For example \cite{pedebearn2024b} suggests including abstracted non-border network nodes in the topology aggregation exposed by network domains. A potential application is the creation of a multi-domain Federated Mission Network \cite{hassan2023multi} between ally nations on top of a partially shared network infrastructure. In our work, we assume that within a network coalition, each participant is ready to share up-to-date information on its network infrastructure (be it in the form of a detailed view or a crude abstraction) with the other ones. 

Another sizeable challenge posed by the multi-domain orchestration of virtualized network services is related to security issues. Indeed, by steering network traffic to virtualized network functions deployed in third-party-operated infrastructures, NFV clearly opens up Confidentiality, Integrity, and Availability vulnerabilities~\cite{Forti2020}. Hence, some critical network functions may have to be placed in trusted infrastructures. For instance, to comply with privacy legislation, a company may have to process user data at a specific private location under its control. Similar security concerns obviously arise in the context of federated military networks. The virtual network embedding problem under security constraints has been addressed in \cite{Alaluna2020,Wang2015}. More recently, \cite{Torkzaban2019} and \cite{Torkzaban2020} introduce the notion of trustworthiness as a measure of security in network service deployment so as to ensure that VNFs performing mission critical operations are hosted on a trusted infrastructure. However, they focus more on the availability of network services (e.g. resilience to attacks) than on confidentiality and integrity issues.  In our work, we adopt a different approach and assume that a network service may only be  deployed within a coalition of networks with mutual trust. The trust relationships between network operators are given as inputs to our problem and may be specific to each network service, as the confidentiality and integrity requirements are not the same for critical services than for standard services.

As mentioned above, we also consider the \gls{VNF-FGE} problem in an online setting where requests for new logical networks arrive one after the other in time.  Each logical network has to be deployed (or rejected) immediately upon arrival. Exact or approximate algorithms for the placement of each individual logical network have been proposed in \cite{taghavian2023fair,esfandyari2025online,Mohamed2022}. Note however that, as individual placement and routing decisions are made without knowing neither the future requests nor the logical networks that will be removed in the future, they are necessarily sub-optimal in hindsight. \added{The authors of \cite{el2022game} and \cite{quang2018single} suggest a periodic reallocation of network slices so as to maintain a close-to-optimal solution over time and propose polynomial-time heuristic algorithms for that.}
\added{Similarly, the authors in \cite{johari2023drl} and \cite{liu2024reinforcement} periodically re-optimize slice embeddings using DQN-based reinforcement learning to improve slice embedding and end-to-end resource allocation in dynamic network slicing environments, but they operate in single-operator settings ignoring trust constraints and offer no optimality benchmark.
In \cite{esfandyari2025online}, the authors leverage actor-critic reinforcement learning and genetic algorithms for online VNF placement in 5G networks; however, the work assumes a single administrative domain and does not consider multi-operator environments or trust constraints in service function chain embedding.}
In our work, we formulate the placement of an individual  logical network as an \gls{ILP} problem. Our approach is based on a path-link formulation handling the above-mentioned trust constraints  and experiments show that this formulation can be solved to optimality for fairly large networks. We also introduce a dynamic pricing scheme for network resources to incentivize efficient resource allocation over the time by slice providers, thereby mitigating the need for periodic reallocation of network slices.

\section{Network Model}
\label{Network Model}
We present below our network model. The model of the \gls{PSN} is described in Section \ref{Model of the Physical Network}, whereas Section \ref{Virtual Network Functions} and \ref{Network Slice Model} are devoted to VNF and network slice models, respectively. Finally, Section \ref{Trust Relationship} presents our model of trust relationship between network operators. \added{Our notation is summarized in Table \ref{tab:notation}.}

\begin{table*}[t]
\color{black}
    \caption{Notation \label{tab:notation}}
    \begin{tabularx}{\textwidth}{p{0.18\textwidth}X}
    \toprule
      {\underline{\bf Physical Network:} } \\
      $\cG$ (resp. $\cG_i$) & physical network (resp. of operator $i=1,\ldots,N$)\\
      $\cD$ (resp. $\cD_i$) & set of function nodes (resp. in network $i=1,\ldots,N$)\\
      $c_e,\tau_e,\phi_e$ & capacity, propagation delay and cost per unit capacity of link $e$ \\
      $\Vec{c}_v$ (resp. $\Vec{\psi}_v$) &  Vector of residual capacities (resp. costs/unit) for each resource $r \in \cR$ at function node $v$\\
      & \\
      {\underline{\bf VNFs:} } \\
      $\cF$ (resp. $\cF^\cS$)  & set of VNFs (resp. required by network slice $\cS$)\\
      $\gamma_f$   & per-packet processing delay of VNF $f$  \\
      $\Vec{p}_f$   & Vector of required amounts of each resource type $r \in \cR$ for executing VNF $f$\\
    & \\      
      {\underline{\bf Network Slice:} } \\
      $(V^g;E^g)$, $b^g$, $\tau_{max}^g$ &  logical service graph for network service $g \in \cS$, its bandwidth requirement between nodes $s^g$ and $t^g$ and its maximum end-to-end latency\\
      $\lambda_f$, $\cD(f)$ & total bandwidth required for VNF $f \in \cF^\cS$ and function nodes where it can be deployed\\
    & \\      
      {\underline{\bf Trust Relationship:} } \\
      $T$  & binary relation such that $i T j$ if and only if network $\cG_i$ trusts network $\cG_j$\\      
      \bottomrule
     \end{tabularx}
\color{black}
\end{table*}

\subsection{Model of the Physical Network}
\label{Model of the Physical Network}
We consider a physical network $\cG = (\cV, \cE)$, which is an interconnection of $N$ individual networks $\cG_i$ = ($\cV_i$, $\cE_i$), where $\cV_i$ is the set of nodes and $\cE_i$ is the set of links of network $i = 1, 2, \hdots ,N$. Each individual network is operated by a distinct network provider and belongs to a specific network domain (e.g., RAN, Edge, Transport and Core networks).
We distinguish between two main types of resources:
\begin{itemize}
    \item \textbf{Communication links:} each communication link $e\in \cE$ is characterized by its residual capacity $c_e$, its propagation delay $\tau_e$ and its cost per unit capacity $\phi_e$.
    \item \textbf{Function nodes:} some of the nodes in $\cV$ have the required compute, storage and networking resources to run \glspl{VNF}. Those nodes are called function nodes and we denote by $\cD_i\in\cV_i$ the set of function nodes in network $\cG_i$, and by $\cD = \bigcup^N_{i=1}\cD_i$ the set of all function nodes. The other nodes are regarded as forwarding nodes (i.e., switches and routers). The execution of a VNF requires different resource types (e.g., CPU, Storage and RAM). We let $\cR$ be the set of those resource types and represent by the vector $\Vec{c}_v=(c^r_v)_{r\in\cR}$ the amount of resources of each type available at node $v\in\cD$. We let $\Vec{\psi}_v = (\psi_v^r)_{r\in\cR}$, where $\psi_v^r$ is the cost of one unit of capacity of resource $r$ at node $v$.
\end{itemize}

\subsection{Virtual Network Functions}
\label{Virtual Network Functions}
As mentioned above, the function nodes are able to host \glspl{VNF}. We let $\cF$ be the set of \glspl{VNF}. The function $f$ introduces a per-packet processing delay $\gamma_f$. The vector $\Vec{p}_f=(p^r_f)_{r\in\cR}$ represents the amount of resource of each type required by \gls{VNF} $f$ for processing one unit of traffic. We assume that this vector is independent of the function node where the \gls{VNF} is deployed.
\begin{figure}[!t]
	\begin{centering}
		\includegraphics[trim={0cm 0cm 0cm 0cm},width=0.5\textwidth]{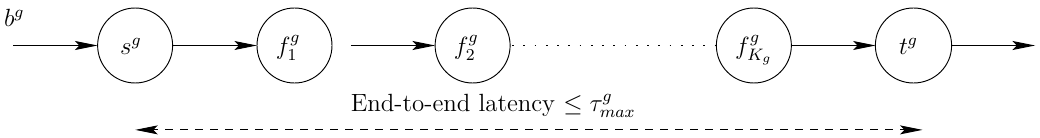}
		\caption{Notations used to describe a network service.}
		\label{fig:VNF-FG}
	\end{centering}
\end{figure}

\subsection{Network Slice Model}
\label{Network Slice Model}
We consider a network slice $\cS$ which must be mapped to existing network capabilities of the physical network. We let $\cF^{\cS}$ be the set of \glspl{VNF} associated to network slice $\cS$. The network slice is defined as a collection of \glspl{VNF-FG} forming end-to-end service. 
Each network service $g\in \cS$ is deployed between two specific locations $s^g$ (service ingress point) and $t^g$ (service egress point) and is an interconnection of \glspl{VNF} providing a composite functionality/behaviour.
As illustrated in Fig. \replaced{\ref{fig:VNF-FG}}{2}, a graph $(V^g;E^g)$ is associated to each network service $g\in \cS$.
For simplicity, we assume that this graph is a linear graph, that is, the packets using the service have to go through a sequence of \glspl{VNF} in a certain prescribed order. The nodes of this graph are either the service ingress/egress points or the \glspl{VNF}, and the edges represent the Virtual Links established between the \glspl{VNF}. We denote by $b^g$ the required bandwidth between the source node $s^g$ and the destination node $t^g$, and by $\lambda_f=\sum_{g\in\cS:f\in V^g}b^g$ the total bandwidth required for VNF $f\in \cF^{\cS}$. We also denote by $\tau^g_{max}$ the end-to-end maximum latency for the network service $g$.
\par
Some \glspl{VNF} of the network slice $\cS$ can be deployed on any function node, while others can only be deployed only on certain function nodes for security or latency reasons. We let $\cD(f)\subseteq \cD$ be the set of function nodes where \gls{VNF} $f\in \cF^\cS$ can be deployed.

\subsection{Trust Relationship}
\label{Trust Relationship}
\added{In addition to the above constraints, when a network slice is deployed across multiple domains, the slice deployment should also take care of the trust relationships between network operators, as we now explain.}
As user requests are made at the edges of the network and often require processing
in the network cores, collaboration between the different network operators is necessary. This implies that network operators have to share information not only about the network slice $\cS$ to be deployed, but also information about their own topologies. Obviously, some operators may be reluctant to share such information with some others. In other words, the information sharing required for establishing a network slice should respect a certain trust relationship between network operators. We shall model this trust relationship as a \added{reflexive and symmetric} homogeneous binary relation $T$ over the set $[1,N]$. We write $iTj$ to indicate that the network $\cG_i$ trusts the network $\cG_j$. The trust relationship may be represented by a graph whose set of nodes are the integers $i\in[1,N]$, and such that there is an edge between nodes $i$ and $j$ if and only if $iTj$.
\par
\begin{figure}[t]
    \centering
    \begin{subfigure}{0.35\textwidth}
        \centering
        \includegraphics[width=0.9\linewidth]{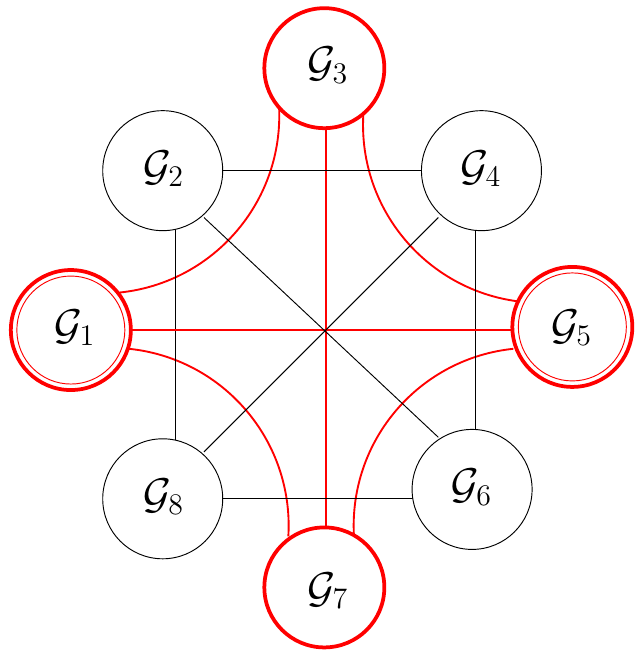}
        \caption{Transitive relationship.}
        \label{fig:Transitive trust}
    \end{subfigure}
    \hfill
    \begin{subfigure}{0.35\textwidth}
        \centering
        \includegraphics[width=0.9\linewidth]{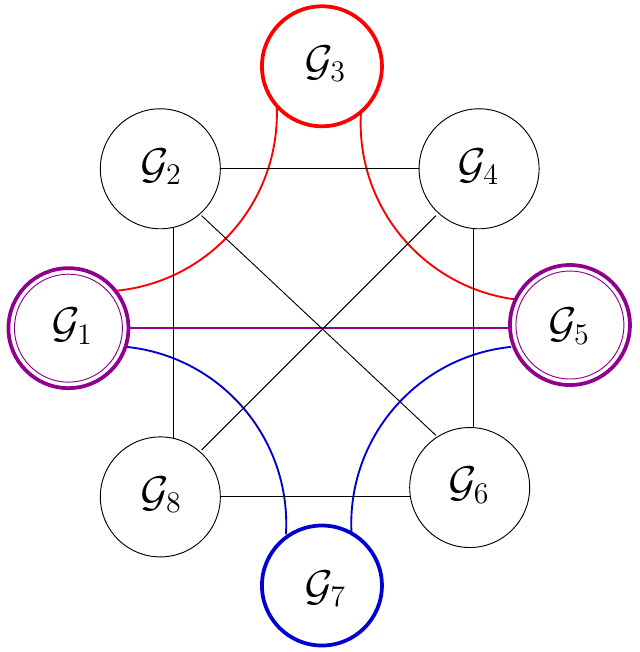}
        \caption{Non-transitive relationship.}
        \label{fig:Non-transitive trust}
    \end{subfigure}

    \caption{Illustration of transitive and non-transitive trust relationships. The source and destination networks are $\cG_1$ and $\cG_5$, respectively. Fig. \ref{fig:Transitive trust} illustrates the case of a transitive trust relationship for which there are two equivalence classes. The equivalence class of the source network $\cG_1$ is shown in red. Fig. \ref{fig:Non-transitive trust} illustrates the case of a non-transitive trust relationship. In this case, the source and destination networks belong to two different coalitions, $\{\cG_1, \cG_3, \cG_5\}$ and $\{\cG_1, \cG_7, \cG_5\}$.}
    \label{fig:main}
\end{figure}

For any \gls{VNF-FG} $g\in\cS$, we shall assume that if $s^g\in\cG_i$ and $t^g\in\cG_j$, then $iTj$, since otherwise it is not possible to map the \gls{VNF-FG} onto the physical network while satisfying the trust relationship. We consider two different types of trust relationship:
\begin{itemize}
    \item \textbf{Transitive trust relationship,} that is, trust relationships $T$ such that $iTj$ and $jTk$ implies $iTk$. In this case, the trust relationship is an equivalence relation and the source and destination networks are in the same equivalence class for the relation $T$. As illustrated in Fig. \replaced{\ref{fig:Transitive trust}}{3a}, this equivalence class corresponds to a maximal clique containing the source and destination networks in the graph associated to the trust relationship. This equivalence class can be interpreted as a coalition of networks with mutual trust. All networks not belonging to this coalition could be pruned from the graph $\cG$ as they are not trusted by the source and destination networks.
    \item \textbf{Non-transitive trust relationship,} in which case the source and destination networks may belong to multiple distinct maximal cliques, or, to say it differently, to multiple coalitions of networks. This situation, which is illustrated in Fig. \replaced{\ref{fig:Non-transitive trust}}{3b}, is more complex as we have to consider the deployment of the network slice in each coalition of networks.
\end{itemize}

\section{Node-Link Formulation of the Online Embedding Problem}
\label{Node-Link Formulation}
We now consider the VNF placement and chaining problem for an incoming network slice request. We are given a network slice $\cS$ and the goal is to find a minimum-cost placement of the \glspl{VNF} $f\in\cF^\cS$ as well as placement of the virtual links required by the network services $g\in\cS$ onto the physical network $\cG$. The state of the \gls{PSN} is assumed to be known, that is, the residual capacities $c_e$ and $\Vec{c}_v$ are known for all links $e \in \cE$ and all function nodes $v\in\cV$. The mapping should satisfy delay and capacity constraints, as well as the placement constraints for the \glspl{VNF}. The trust relationship between network operators should also be respected. The problem is formulated below as an integer linear program. \added{The formulation uses the binary variables $x_{v,f}$, $y_l^{e,g}$ and $z^g_v$, which are as defined in Table \ref{tab:notation-pb}.} \added{We assume that the goal is to minimize the total cost of the \gls{VNF} placement and chaining decisions, which can be expressed as follows:
    \begin{equation}
        \sum_{e\in\cE}\phi_e \sum_{g\in\cS}\sum_{l\in E^g}y^{e,g}_lb^g+ \sum_{v\in\cD}\sum_{r\in\cR}\psi^r_v \sum_{f\in V^\cS}x_{v,f}\lambda_f p^r_f
    \end{equation}
}
\noindent \added{where $\phi_e$ is the cost per unit capacity of link $e$, and $\psi^r_v$ is the cost of unit capacity of resource $r$ at node $v$.}

\begin{table*}[t]
\color{black}
    \caption{Variables for Node-Link and Path-Link Formulations \label{tab:notation-pb}}
    \begin{tabularx}{\textwidth}{p{0.15\textwidth}X}
    \toprule
      {\underline{\bf Node-Link:} } \\
      $x_{v,f}$ & binary variable which is $1$ if VNF $f$ is deployed at function node $v\in\cD(f)$, and 0 otherwise.\\
      $y_l^{e,g}$ & Given a virtual link $l = (f, f^\prime)\in E^g$, this binary variable indicates whether link $e\in\cE$ is on the path connecting the VNFs $f$ and $f^\prime$.\\
      $z^g_v$ & binary variable which is $1$ if node $v$ is used to map network service $g\in\cS$, and 0 otherwise\\
      & \\
      {\underline{\bf Path-Link:} } \\
      $x_\pi^g$ & binary variable which is $1$ if candidate path $\pi \in \Pi^g$ is retained for deploying service $g \in \cS$\\
      $y_e$   & traffic flowing on the physical link $e$  \\
      $y_{v,f}$   & total traffic processed by VNF $f$ at node $v \in \cD(f)$\\
      \bottomrule
     \end{tabularx}
\color{black}
\end{table*}

\added{The optimization is made subject to the following constraints:}

\begin{itemize}
    \item \textbf{Placement constraints:} \added{the constraint}
    \begin{equation}
        \label{eq:placement constraint}
        \sum_{v\in\cD(f)} x_{v,f}=1, \mbox{for all $f\in\cF^\cS$,}
    \end{equation}
    expresses that each \gls{VNF} $f\in\cF^\cS$ has to be mapped to one and only one node $v\in\cD(f)$.
    \item \textbf{Routing constraints:} if $l\in E^g$, then we need to find a path in $\cG$ connecting the nodes where the \glspl{VNF} corresponding to the endpoints of $l$ are placed (provided these nodes are different). This path is a segment of the whole path connecting $s^g$ and $t^g$ and will be used to setup a Virtual Link connecting the \glspl{VNF}. To model this \added{we use the binary variable $y_l^{e,g}$ (see Table \ref{tab:notation-pb})}. We remind the reader that \gls{VNF} $f$ is placed on node $v$ if and only if $x_{v,f} = 1$, and similarly \gls{VNF} $f^\prime$ is placed on node $v^\prime$ if and only if $x_{v^\prime,f^\prime} = 1$. Hence, the existence of a path connecting the nodes where those \glspl{VNF} are placed can be enforced with the following flow conservation constraints:
    \begin{equation}
    \label{eq:routing constraints}
    \sum_{e\in O(n)} y_l^{e,g} - \sum_{e\in I(n)} y_l^{e,g} =
    \begin{cases}
    \begin{aligned}
        x^g_{v,f} & \quad \text{if } n = v, \\
        -x^g_{v^\prime,f^\prime} & \quad \text{if } n = v^\prime, \\
        x^g_{v,f} - x^g_{v^\prime,f^\prime} & \quad \text{if } n = v \\
        & \quad \text{and } n = v^\prime, \\
        0 & \quad \text{otherwise.}
    \end{aligned}
    \end{cases}
    \end{equation}
    for all nodes $n\in\cV$, for all nodes $v\in\cD(f)$ and $v^\prime\in\cD(f^\prime)$ such that $v^\prime \neq v$, all virtual links $l= (f, f^\prime)\in E^g$, and all network service $g\in\cS$. Note that in equation \eqref{eq:routing constraints}, $O(n)$ (resp. $I(n)$) is defined as the set of links outgoing from (resp. incoming to) node $n$.
    \item \textbf{Trust constraints:} in order to model the trust constraints, we use the binary variable $z^g_v$ (see Table \ref{tab:notation-pb}). We impose the equality $z^g_v = 1$ for $v\in \{s^g,t^g\}$. For the other nodes $v\in\cV\backslash \{s^g,t^g\}$, the value of $z^g_v$ is defined by the following constraints:
    \begin{align}
        \label{eq:trust constraints 1}
        M\space z^g_v &\geq \sum_{e\in I(v)}\sum_{l\in E^g} y^{e,g}_l, \\
        z^g_v &\leq \sum_{e\in O(v)}\sum_{l\in E^g} y^{e,g}_l,
        \label{eq:trust constraints 2}
    \end{align}

    where $M$ is a large constant. The RHS of \eqref{eq:trust constraints 1} (resp. \eqref{eq:trust constraints 2}) represents the number of virtual links of network service $g$ using a physical link that is incoming to (resp. outgoing from) node $v$. \added{Consider now two networks $\cG_i$ and $\cG_j$ such that there is no trust between them. If node $v\in\cG_i$ hosts a VNF for service $g$, no node $v^\prime\in\cG_j$ can host a VNF for the same service, and vice-versa. This trust relationship is enforced by imposing the following constraint }

    \begin{equation}
    \label{eq:trust constraints 3}
        z^g_v + z^g_{v^\prime} \leq 1, \quad \mbox{for all } v\in\cG_i \mbox{ and }  v^\prime\in\cG_j.
    \end{equation}
    
    \item \textbf{Latency constraints:} the end-to-end delay of a network service $g\in\cS$ is composed of a fixed processing delay which is $\sum_{f\in V^g}\gamma_f$ and it should not exceed $\tau^g_{\text{max}}$. It yields the following constraint for the sum of the delays incurred over the edges:
    \begin{equation}
        \sum_{e\in\cE}\sum_{l\in E^g} y^{e,g}_l \tau_e\leq \tau^g_{\text{max}} - \sum_{f\in V^g}\gamma^f, \text{ for all } g\in\cS.
    \end{equation}
    \item \textbf{Capacity constraints:} The capacity constraint on the links can be enforced by imposing that
    \begin{equation}
        \sum_{g\in\cS}\sum_{l\in E^g} y^{e,g}_lb^g \leq c_e, \text{ for all }e\in\cE.
    \end{equation}

    Regarding the capacities of the function nodes, we impose the following constraints:
   \begin{equation}
        \sum_{f:v \in\cD(f)} x_{v,f} \lambda_f p_f^r \leq c_v^r, \text{ for all }v\in\cD, r \in \cR.
    \label{eq:cstr-capacity-nodes}
    \end{equation}

\end{itemize}
\par
The above problem is hard to solve, and thus requires high computational time to find the optimal solution.
In the next section, we formulate an alternative approach to reduce the time complexity of the problem.

\section{Path-Link Formulation of the Online Embedding Problem}
\label{Path-Link Formulation}
In this section, we propose a PL formulation of the VNF placement and chaining problem discussed in Section \ref{Node-Link Formulation}. In contrast to the NL formulation discussed there, which implicitly takes into account all possible allocations, the PL formulation makes direct use of a predefined list of candidate allocations, which correspond to paths in an expanded physical network. The main advantage is that a reduced computational time can be expected. The downside is of course that the optimality of the solution can no more be guaranteed. We first introduce the concept of expanded network in Section \ref{Expanded Network}. We then discuss the generation of candidate paths satisfying the trust relationship in Section \ref{Trusted Networks and Candidate Paths}, before delving into the details of the path-link formulation in Section \ref{Formulation as a Single-Path Routing Problem}.

\subsection{Expanded Network}
\label{Expanded Network}
Consider a network service $g \in \mathcal{S}$, for which the \glspl{VNF} $f \in \mathcal{F}^g$ must be placed on function nodes and traversed in a prescribed order. The end-to-end path from $s^g$ to $t^g$ can therefore be viewed as a sequence of path segments interleaved with decisions on where each VNF is executed.
\par
Following \cite{nguyen2017optimizing}, this joint placement-and-routing problem can be modeled using an expanded network. For a given \gls{VNF-FG} $g$, the expanded network is a layered directed graph composed of $k^g+1$ copies of the physical network $\mathcal{G} = (\mathcal{V}, \mathcal{E})$. Inter-layer artificial edges represent the execution of VNFs at admissible function nodes, while intra-layer edges correspond to physical links. Artificial edges are weighted by VNF processing delays $\gamma_f$, whereas physical edges are weighted by link propagation delays $\tau_e$.
\par
\begin{figure}[!t]
	\begin{centering}
		\includegraphics[trim={0cm 0cm 0cm 0cm},width=0.35\textwidth]{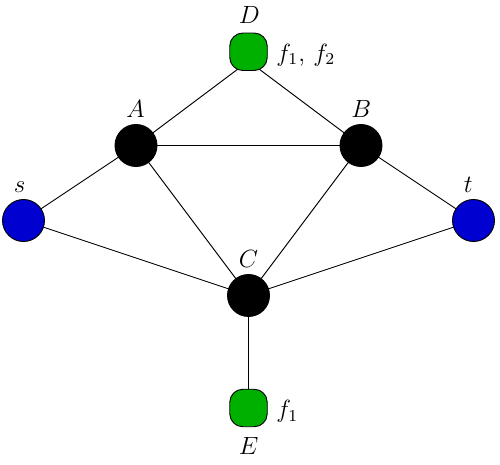}
		\caption{Example network in which the source node is $s$, the destination node is $t$, and VNFs $f_1$ and $f_2$ have to be visited in that order. The VNF $f_1$ can be hosted by nodes $D$ and $E$, whereas $f_2$ is only available at node $D$.}
		\label{fig:example-network}
	\end{centering}
\end{figure}
\begin{figure}[!t]
	\begin{centering}
		\includegraphics[trim={0cm 0cm 0cm 0cm},width=0.79\textwidth]{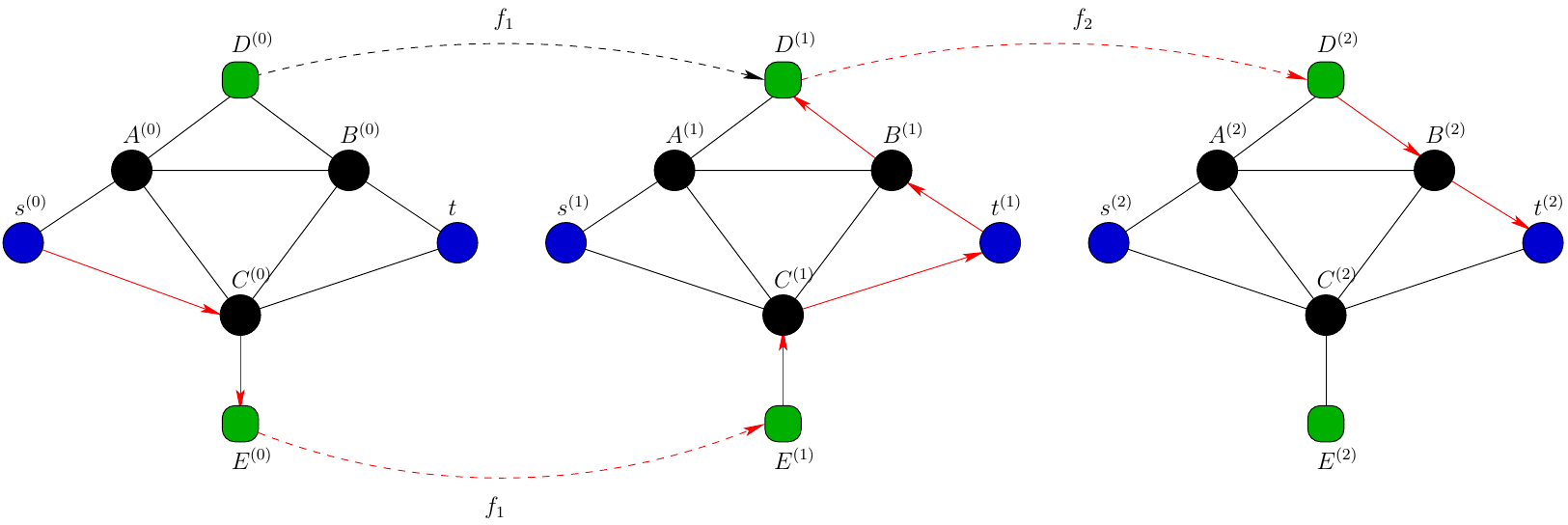}
		\caption{Expanded network for the example of Fig. \ref{fig:example-network}. The artificial edges between consecutive layers are shown with dotted lines. A feasible path from node $s$ at layer 0 to node $t$ at layer 2 is shown in red.}
		\label{fig:expanded-network}
	\end{centering}
\end{figure}

We illustrate the expanded network construction using the example in Fig.~\ref{fig:example-network}, where a \gls{VNF-FG} connects source $s$ to destination $t$ and requires \glspl{VNF} $f_1$ and $f_2$ to be executed in order. Function $f_1$ can be hosted at nodes $D$ or $E$, while $f_2$ is only available at node $D$. The corresponding expanded network, shown in Fig.~\ref{fig:expanded-network}, consists of three layers connected by artificial edges.
\par
Any path $\pi$ from the copy of $s^g$ in layer~0 to the copy of $t^g$ in layer $k^g\!-\!1$ represents a feasible VNF placement and chaining solution: artificial edges indicate where VNFs are executed, while physical edges define the routing between them, enforcing the prescribed order. Fig.~\ref{fig:expanded-network} highlights such a path.
\par
In the following, candidate solutions are obtained by enumerating paths $\pi$ in the expanded network. The length $\tau(\pi)$ represents the end-to-end delay, and only paths satisfying $\tau(\pi) \leq \tau^g_{\max}$ are considered feasible.

\subsection{Trusted Networks and Candidate Paths}
\label{Trusted Networks and Candidate Paths}
As shown in Section \ref{Expanded Network}, a solution to the \gls{VNF} placement and chaining problem for a \gls{VNF-FG} $g \in \cS$ can be represented as a path in the corresponding expanded graph. By construction of the expanded graph, the path specifies a solution in which each VNF $f \in \cF^g$ is executed at a function node $v \in \cD(f)$, so that location constraints on the deployment of the VNFs are satisfied. Moreover, it is also guaranteed that the \glspl{VNF} are executed in the prescribed order. As discussed above, it is also fairly easy to enforce the end-to-end latency constraint by considering only paths $\pi$ such that $\tau(\pi) \leq \tau_{max}^g$. We now explain how to ensure that those paths also satisfy the trust relationship.
\par
As discussed in Section \ref{Trust Relationship}, for a transitive trust relationship, the source network $\cG_i$ and the destination network $\cG_j$ belong to a unique network coalition, which is an equivalence class of the trust relation $T$. In that case, we can obtain candidate paths respecting the trust relationship by simply removing all networks $\cG_k$ not belonging to this equivalence class, and then choosing some candidate paths in the expanded graph of the residual networks between nodes $s^g$ and $t^g$. These paths can be chosen for instance by solving a $K$-shortest path problem in the expanded graph. 
\par
For a non-transitive trust relation $T$, the situation is more complex as $\cG_i$ and $\cG_j$ can belong to multiple network coalitions. A network coalition is a subset $\cA \subseteq \{1,\ldots,N\}$ such that $x \in \cA$ and $y \in \cA$ implies $xTy$. We call such a network coalition a \emph{trusted network}. For a non-transitive trust relation, the idea is simply to consider all possible trusted networks to which $\cG_i$ and $\cG_j$ belong, and to generate some candidate paths in the expanded graph of each one. Given a trusted network, the choice of the candidate paths can again be done using a $K$-shortest path algorithm. 
\par
A trusted network corresponds to a maximal clique (that is, a complete subgraph) in the trust graph, which is an undirected graph whose set of vertices is $\{1,\ldots,N\}$, and such that there is an edge between vertices $i$ and $j$ if and only if $iTj$. We therefore have to list all maximal cliques containing the nodes $i$ and $j$ in the trust graph. Finding all maximal cliques in a given graph is a fundamental problem in graph theory, known as the Maximal Clique Enumeration (MCE) problem. It is one of Karp’s 21 NP-complete problems. We note that the lack of a polynomial-time algorithm for solving this problem is not an issue in our case, as we shall only consider "small" trust graphs. \added{Indeed, it is important to keep in mind that the nodes $k$ in the trust graph correspond to the individual networks $\cG_k$ in the original problem, so they have at most a few tens of nodes}.
\par
In the literature, the MCE problem has been widely studied. One of the most notable, efficient, successful and extensively used solutions is the Bron–Kerbosch (BK) algorithm. It is a recursive algorithm which is able to enumerate all maximal cliques in an undirected graph without duplication. \added{Its worst-case running time is $O(3^{n/3})$ and numerical experiments have demonstrated that it runs very fast in practice~\cite{Tomita2006}. In particular, on average the algorithm takes less than 2\% of the total simulation runtime.} We use the basic version of this algorithm and describe below how it can be used to generate candidate paths satisfying the trust relationship.
\par
The Bron–Kerbosch algorithm is a recursive backtracking algorithm, whose pseudocode is given in Algorithm \ref{algo:bron-kerbosch}. It takes as input three disjoint sets of vertices $R$, $P$, and $X$ of the trust graph and finds the maximal cliques that include all of the vertices in $R$, some of the vertices in $P$, and none of the vertices in $X$. Assuming that the source and destination networks are $\cG_i$ and $\cG_j$, the recursion is initiated in our case by setting $R=\{i,j\}$, $P=\left \{1,\ldots,N\right \} \setminus \{i,j\}$ and $X=\emptyset$. 

\begin{algorithm}
	\caption{Bron-Kerbosch Algorithm}
	\label{algo:bron-kerbosch}
	\begin{algorithmic}[1]
		\Require $R$, $P$, $X$
		\If{$P=\emptyset$ and $X=\emptyset$}
			\State {Compute $K$ shortest path in the expanded graph of $\bigcup_{k \in R} \cG_k$} 
		\EndIf
		\For {each vertex $v \in P$}
		    \State Call BronKerbosch($R \cup \{v\}$, $P \cap N(v)$, $X \cap N(v)$)
		  \State $P \leftarrow P \setminus \{v\}$
		  \State $X \leftarrow X \cup \{v\}$
		\EndFor
	\end{algorithmic}
\end{algorithm}
\par
In each call to the algorithm, $P$ and $X$ are disjoint sets whose union consists of those vertices that form cliques when added to $R$. When both $P$ and $X$ are empty, $R$ is a maximal clique and we can compute some candidate paths in the expanded graph of the residual network composed of the networks $\left \{ \cG_k \right \}_{k \in R}$. More precisely, the candidate paths are computed in the expanded graph of the induced subgraph corresponding to the set of nodes $\bigcup_{k \in R} \cV_k$. If $X$ is empty but $P$ is not, then the algorithm makes a recursive call for each vertex $v \in P$ to find all clique extensions of $R$ that contain $v$. In this recursive call, $v$ is added to $R$ and $P$ and $X$ are restricted to the neighbour set $N(v)$ of $v$.

\subsection{Formulation as a Single-Path Routing Problem}
\label{Formulation as a Single-Path Routing Problem}
In what follows, we assume that for each \gls{VNF-FG} $g\in\cS$, we are given a set $\Pi^g$ of candidate paths in the corresponding expanded network. Those paths are obtained as described in Sections \ref{Expanded Network} and \ref{Trusted Networks and Candidate Paths}. Each one specifies a possible \gls{VNF} placement and chaining solution for a network service $g$, such that: (a) the \glspl{VNF} are executed in the prescribed order, (b) the placement constraint of each \gls{VNF} is satisfied, (c) the end-to-end latency constraint is satisfied, and (d) the trust constraints are respected. A path in $\Pi^g$ also specifies the path
segments to be used to interconnect consecutive \glspl{VNF} in the sequence $\cF^g$.
\par
\added{We use the decision variables defined in Table \ref{tab:notation-pb}.} A solution of the \gls{VNF} placement and chaining problem is defined as a vector $\pi=(\pi^g)_{g\in\cS}$. Given such a solution, the traffic $y_e$ flowing on the link $e\in\cE$ of the original network can be expressed as:
\begin{equation}
    y_e = \sum_{g\in\cS}\theta^e_{\pi^g}b^g,
    \label{eq:def-ye}
\end{equation}
where the constant $\theta^e_\pi$ is defined as the multiplicity of edge $e$ in path $\pi$ (it may appear at multiple layers of the expanded network). Similarly, we can write the total traffic processed by \gls{VNF} $f$ at node $v\in\cD(f)$ as follows:
\begin{equation}
    y_{v,f} = \sum_{g\in\cS}\theta^{v,f}_{\pi^g}b^g.
    \label{eq:def-yvf}
\end{equation}
Here, the constant $\theta^{v,f}_\pi$ represents the number of times that an artificial edge associated to \gls{VNF} $f$ and connecting two consecutive copies of node $v$ appears in path $\pi$. Note that we have $\theta^{v,f}_{\pi^g}=0$ if $f \not\in \cF^g$, of if $v\not\in\cD (f)$. The constant $\theta^{v,f}_\pi$ may be greater than 1 if the \gls{VNF} $f$ appears multiple times in the sequence $\cF^g$.
\par
A solution $\mathbf{\pi}$ is feasible if the capacity constraints of the links and function nodes are satisfied, that is, $y_e\leq c_e$ for all $e\in\cE$ and $\sum_{f\in\cF}y_{v,f}\Vec{p}_f\leq \Vec{c}_v$ for all function nodes $v\in\cD$.
Among the set of feasible solutions, we look for a solution minimizing the linear cost:
\begin{equation}
    \sum_{e\in\cE}\phi_e y_e + \sum_{v\in\cD}\sum_{r\in\cR}\psi^r_v y_{v,f} p^r_f
\end{equation}

Define the decision variable $x^g_\pi$ as $1$ if candidate path $\pi \in \Pi^g$ represents the retained solution for deploying network service $g \in \cS$, and $0$ otherwise. Provided that a sufficiently large set of candidate paths is considered for each network service, a near-optimal solution of the online \gls{VNF} placement and chaining problem can be obtained as an optimal solution of the following Mixed Integer Linear Program (MILP):
\color{black}
\begin{align}
    \mbox{min}\; & F(\mathbf{\pi}) = \sum_{e\in\cE}\phi_e y_e + \sum_{v\in\cD}\sum_{r\in\cR}\psi^r_v\sum_{f\in\cF}y_{v,f} p_f^r \\
    \mbox{s.t.: }\; &\nonumber \\
    & y_e = \sum_{g\in\cS}\sum_{\pi\in\Pi^g}\theta^e_\pi b^gx^g_\pi, \mbox{ for all } e\in\cE \label{cstr:pl1} \\
    & y_{v,f} = \sum_{g\in\cS}\! \sum_{\pi\in\Pi^g}\theta^{v,f}_\pi b^g x^g_\pi, \mbox{ for } v \! \in \! \cD(f), f \! \in \! \cF, \label{cstr:pl2} \\
    & y_e\leq c_e, \mbox{ for all } e\in\cE, \label{cstr:pl3} \\
    & \sum_{f\in\cF}y_{v,f}p_f^r\leq c_v^r,  \mbox{ for all } v\in\cD, r \in \cR \label{cstr:pl4} \\
    &y_e \geq 0,  \mbox{ for all } e\in\cE, \label{cstr:pl5} \\
    & y_{v,f} \geq 0,  \mbox{ for all } v\in\cD(f), f\in\cF, \label{cstr:pl6} \\
    & x^g_\pi \in \{0,1\},  \mbox{ for all } \pi\in\Pi^g, g\in\cS \label{cstr:pl7}.
\end{align}
\color{black}
\added{
Constraints \eqref{cstr:pl1} and \eqref{cstr:pl2} define the value of the traffic $y_e$ flowing on link $e$ and of the traffic $y_{v,f}$ processed by VNF $f$ at node $v$, respectively. They directly follow from \eqref{eq:def-ye} and \eqref{eq:def-yvf}. Constraint \eqref{cstr:pl3} is a capacity constraint for link $e$, and similarly, constraint \eqref{cstr:pl4} is a capacity constraint for resource $r$ at node $v$. The other constraints define the values that the decision variables can take.
}

\section{Dynamic Pricing of Network Resources}
\label{Dynamic pricing}
As discussed in sections \ref{Node-Link Formulation} and \ref{Path-Link Formulation}, network slices are placed in the network so as to minimize the total cost of the \gls{VNF} placement and chaining decisions, which has the form $\sum_{e}\phi_e y_e + \sum_{v,r}\psi^r_v y^r_v$.
The only way for operators to influence the placement of network slices in their infrastructures is therefore through the costs per unit of capacity $\phi_e$ and $\psi^r_v$ that they publicly advertise. In what follows, we interpret these costs as incentives set by the network providers to drive slice tenants/providers to an efficient resource allocation and explain how to set them so as to optimize the residual capacity of the network.
\par
We assume that to each node $v \in \cV$ and to each resource $r\in\cR$ of the physical network is associated a strictly increasing function $\Phi_v^r : \mathbb{R}_+ \rightarrow \mathbb{R}_+$ and that the cost of this resource has the form $\rho_v^r \Phi_v^r(\rho_v^r)$, where $\rho_v^r$ denotes the utilization rate of that resource. The function $\Phi_v^r()$ may be interpreted as the cost per unit of capacity of resource $r$ at node $v$, and it depends on its utilization rate. Similarly, it is assumed that the cost of link $e\in\cE$ takes the form $\rho_e \Phi_e(\rho_e)$ for some given strictly increasing function $\Phi_e$. A typical example is the Kleinrock function $\Phi_e(\rho_e) =1/(1-\rho_e)$. This function assumes that the total traffic on a resource is smaller than its capacity and that the cost per unit of capacity grows unboundedly as the former approaches the latter. Other examples are the linear function $\Phi_e(\rho_e)=\rho_e$ or the quadratic function $\Phi_e(\rho_e)= \rho_e^2$, which are valid even for $\rho_e \geq 1$. In the above examples, the cost per unit capacity grows as the utilization rate $\rho_r$ of resource $r$ increases, in contrast with linear models in which $\Phi_v^r$ is assumed to be a constant function. This conveys that the cost per unit of capacity for a highly utilized resource is significantly higher than that of a minimally utilized resource. We emphasize that the functions $\Phi_v^r()$ may differ for different types of resources.
\par
The utilization rates of the network resources and the functions $\Phi_e()$ and $\Phi_v^r()$ are private knowledge for the network operators. Nevertheless, they can be linearized so as to compute the costs $\phi_e$ and $\psi^r_v$ which are publicly advertised:
\begin{itemize}
    \item The cost per unit capacity $\phi_e$ of link $e\in\cE$ is computed as:
    \begin{equation}
        \phi_e=\dfrac{1}{c_e}[\Phi_e(\rho_e) + \rho_e\Phi^\prime_e(\rho_e)]
        \label{eq:cost-per-unit-capacity-from-function-phi}
    \end{equation}
    \item The cost per unit capacity $\psi_v^r$ of resource $r$ at node $v$ is computed as:
    \begin{equation}
        \psi_v^r = \frac{1}{c_v^r} \, \left [ \Phi_v^r(\rho_v^r) + \rho_v^r \, {\Phi_v^r}^\prime(\rho_v^r)\right ] 
        \label{eq:cost-per-unit-capacity-from-function-phi}
    \end{equation}
\end{itemize}
\par
These costs per unit capacity are of course updated each time a new network slice is placed in the network. In that way, it is possible to approximately minimize the total network cost $\sum_e \rho_e \Phi_e(\rho_e)+ \sum_{v,r} \rho_v^r \Phi_v^r(\rho_v^r)$ obtained after multiple myopic placements, thus mitigating the need for periodic reallocations.

\section{Simulation and results}
\label{Simulation and results}
In this section, we evaluate the empirical performance of the proposed Path--Link model described in Section~\ref{Path-Link Formulation}. We first describe the network slice types considered and the procedure for generating random instances in Sections~\ref{subsec:network-slices} and~\ref{subsec:generation-of-random-instances}. Section~\ref{subsec:simulator} then presents the parameters of the developed simulator, and Section~\ref{subsec:numerical-results} reports the numerical results.
\added{The main simulation parameters include the slice arrival rate $\lambda$, slice holding time $\tau$, evaluation horizon $T$, inter-operator link probability $\pi$, candidate path set size $k'$, and resource utilization rate $\rho$, which are defined in Sections~VII-A to VII-C.}

\subsection{Network Slices}
\label{subsec:network-slices}
We consider three types of slices each with different traffic size, delay constraint and trust relationship requirements as shown in Table \ref{tab:slice characteristics}. 
\added{These slices represent heterogeneous service profiles and are not tied to specific applications, enabling a comparative evaluation under diverse operating conditions.}
Rate and delay requirements are drawn from uniform distributions in the interval $[a_i,b_i]$, where $i$ is the slice type. 
We consider three levels of trust relationship. The first level (level 1) ignores the trust constraints in (\ref{eq:trust constraints 1}), (\ref{eq:trust constraints 2}) and (\ref{eq:trust constraints 3}) of the node-link formulation and assumes a clique including all network operators in the path-link formulation. In other words, service demands belonging to this slice type can be deployed anywhere in the network and their traffic can flow through any operator. 
The second and third trust levels impose some trust constraints such that trust level 3 is more restrictive, thus allowing fewer operators to share their topologies to deploy the requested service. We detail the trust relationship between operators based on the trust level in Section \ref{subsec:generation-of-random-instances}.

\begin{table}[!t]
	\centering
	\caption{Slice characteristics.}
	\label{tab:slice characteristics}
	\begin{tabular}{cccc}
		\hline
		\multicolumn{1}{l}{\textbf{Slice Type}} & \multicolumn{1}{l}{\textbf{Rate Req.}} & \multicolumn{1}{l}{\textbf{Delay Req.}} & \multicolumn{1}{l}{\textbf{Trust Level}} \\ \hline
		Slice A & [1,3] & [10,30] & Level 1 \\
		Slice B & [2,5] & [10,20] & Level 2 \\
		Slice C & [3,5] & [1,5]   & Level 3 \\ \hline
	\end{tabular}
\end{table}

\subsection{Generation of Random Instances}
\label{subsec:generation-of-random-instances}
We evaluate the path-link algorithm using real network topologies collected from \emph{The Internet Topology Zoo} \cite{knight2011internet}. We consider four network domains: RAN, Edge, Transport and Core. For each domain, we associate three operators with different topologies selected from \cite{knight2011internet}. For each operator, we fix three source nodes and three destination nodes. 
The original topology is then expanded by adding two function nodes and connecting them to some other nodes. 
We further assume that each network domain can host two types of \glspl{VNF}, and each function node can deploy up to two different \glspl{VNF}. This results in a total of eight different \glspl{VNF} that could be deployed over the network at different locations.
Across all instances, the substrate comprises more than $420$ nodes and $3{,}820$ links, i.e., an ISP-scale topology with hundreds of nodes and links.
\par
The links between operators are generated according to a uniform random distribution with parameter $\pi$, the probability that there is a directed link between two nodes from different operators. The higher $\pi$, the more connected the operators are.
We differentiate between two types of links: intra-connection links (between the nodes of the same operator) and inter-connection links (between the nodes of different operators). We assume that all intra-connection and inter-connection links have uniform capacities, with inter-connection links having higher capacity. We further assume that function nodes belonging to the same domain have similar capacities.
These capacities are generated per random instance to yield various congestion scenarios. We consider four congestion levels: bandwidth-limited, node-limited, both-limited and low congestion as shown in Table \ref{tab:congestion type}.
\added{The congestion scenarios are designed to span low, moderate, and high load regimes in order to evaluate the robustness of trust-aware slice admission and embedding decisions under diverse network stress levels.}

\begin{table}[!t]
	\centering
	\caption{Network congestion types}
	\label{tab:congestion type}
	\begin{tabular}{ccc}
		\hline
		\multicolumn{1}{l}{\textbf{Congestion Type}} & \multicolumn{1}{l}{\textbf{Nodes}} & \multicolumn{1}{l}{\textbf{Links}} \\ \hline
		Bandwidth-limited & 0.5 & 0.85 \\
		Node-limited      & 0.85 & 0.5 \\
		Both-limited      & 0.85 & 0.85 \\
		No congestion     & 0.5 & 0.5 \\ \hline
	\end{tabular}
\end{table}

\par
Once the network topology is set, we generate 50 random instances. A random instance is constructed as follows. For each operator, we randomly choose the types of \glspl{VNF} they can deploy on their function nodes. 
Then, traffic requests are generated according to the slice they belong to. The source and destination nodes for each request are randomly selected from their respective sets using a uniform distribution.
The traffic volume and delay requirements for a traffic demand are also drawn from a uniform distribution according to the values set in Table \ref{tab:slice characteristics}. We assume that the traffic volume does not change along the path from source to destination. 
Each traffic demand is also assigned a random processing sequence consisting of a predefined set of \glspl{VNF}.
\par
The final step is to generate the candidate paths for the different traffic requests. These paths must comply with the trust relationship outlined in Section \ref{Trusted Networks and Candidate Paths}. For this, and for each traffic demand, we apply the Bron-Kerbosch algorithm described in Algorithm \ref{algo:bron-kerbosch} in order to find the maximal clique that satisfies the trust relationship.
Once the clique is obtained, each traffic demand must go through several \glspl{VNF} which are available at multiple locations. The possible paths for a traffic demand $g$ are generated as follows.
We first compute two possible paths between the source node $s^g$ and each location hosting the first \gls{VNF} $f_1^g$ by solving a 2-shortest path problem (assuming unit weights for the edges of the expanded network). We then apply the same procedure for computing two possible paths between each possible location of the VNF $f_1^g$ and each possible location of $f_2^g$, etc. The final set of candidate paths $\Pi^g$ is obtained by connecting the different segments forming the overall path from $s^g$ to $t^g$.
\par
The final set of candidate paths for each traffic demand is used to compute the optimal solution using Gurobi solver.
However, when solving for the single-path routing problem, we consider only the $k^\prime$-shortest paths from the previously obtained set.

\subsection{Simulator}
\label{subsec:simulator}
We developed an event-driven stochastic simulator programmed in Python in order to evaluate important network metrics such as service blockage probability and resource utilization rates.
Traffic requests arrive following a Poisson Process with parameter $\lambda$ chosen such that the overall probability of blockage is below 0.2. The connection holding time is generated from an exponential process with an average $\tau = 10$. All requests arriving at a given time slot belong to a single slice type that is randomly chosen among the three different slice types. 
We generate 50 runs, each with a length of $T=200$ time slots. The results are then averaged.

\subsection{Numerical Results}
\label{subsec:numerical-results}
\begin{figure*}[!t]
    \centering

    \begin{subfigure}[b]{0.45\textwidth}
        \centering
        \includegraphics[width=\textwidth]{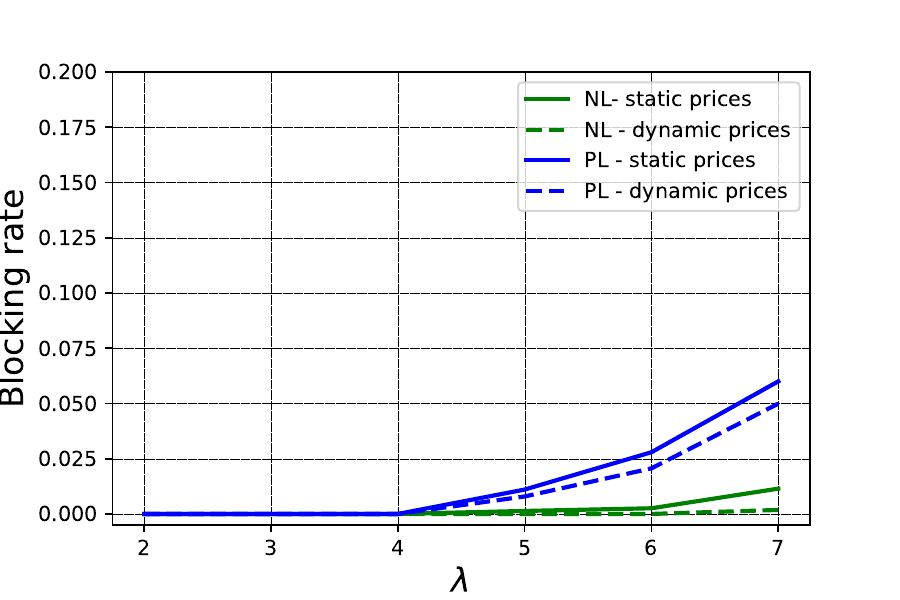}
        \caption{No congestion}
        \label{fig:no congestion}
    \end{subfigure}
    \hfill
    \begin{subfigure}[b]{0.45\textwidth}
        \centering
        \includegraphics[width=\textwidth]{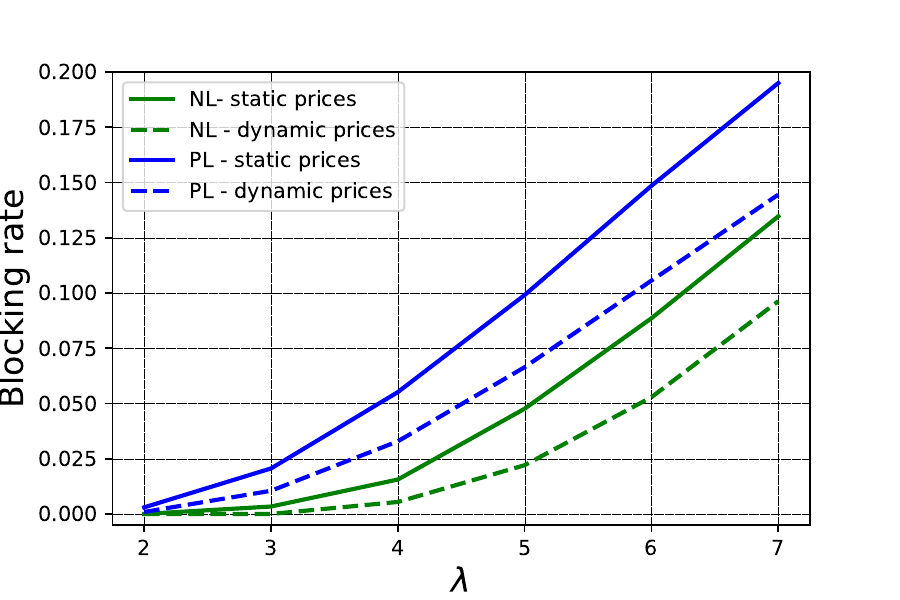}
        \caption{Both limited}
        \label{fig:both limited}
    \end{subfigure}

    \vspace{1em}
    \begin{subfigure}[b]{0.45\textwidth}
        \centering
        \includegraphics[width=\textwidth]{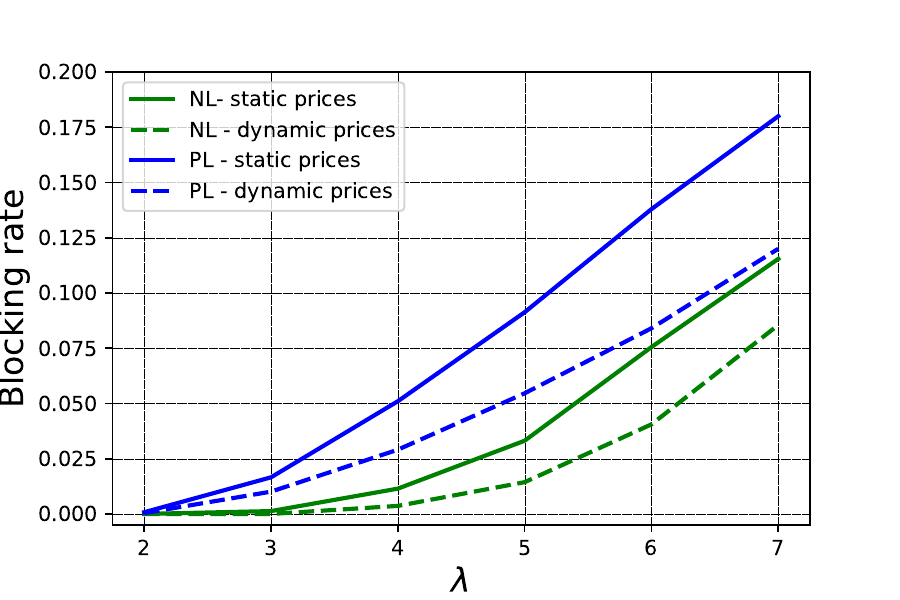}
        \caption{Edge limited}
        \label{fig:edge limited}
    \end{subfigure}
    \hfill
    \begin{subfigure}[b]{0.45\textwidth}
        
        \centering
        \includegraphics[width=\textwidth]{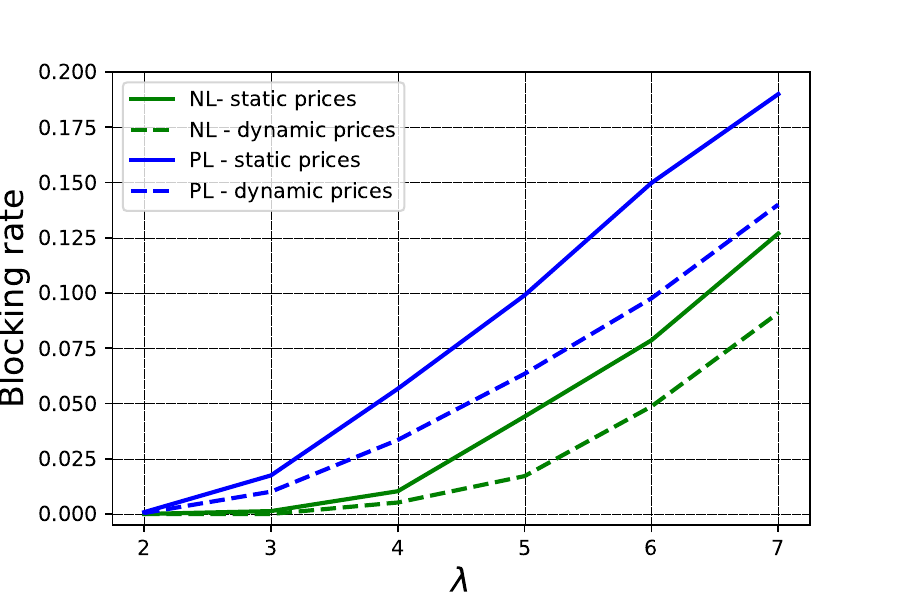}
        \caption{Node limited}
        \label{fig:node limited}
    \end{subfigure}

    \caption{\added{Blocking probability versus arrival rate $\lambda$ under different congestion scenarios.}}
    \label{fig:block rate}
\end{figure*}

\begin{figure}[!t]
	\begin{centering}
		\includegraphics[trim={0cm 0cm 0cm 0cm},width=0.6\textwidth]{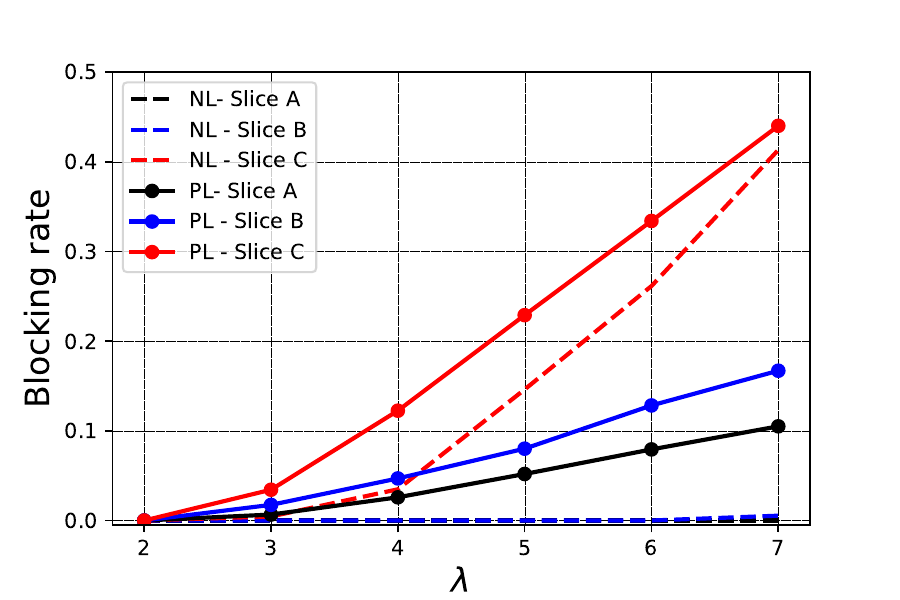}
		\caption{Impact of slice type on the average blocked SFC.}
		\label{fig:Blockage slice}
	\end{centering}
\end{figure}

\added{The plots in Fig.~\ref{fig:block rate}(a-d) provides an overview of the numerical results, illustrating the impact of traffic load, congestion scenarios, dynamic pricing, and the choice between PL and NL formulations on blocking probability.}

\subsubsection{Effect of load}
\added{
As expected, for all congestion scenarios the blocking probability increases monotonically with $\lambda$. Across the entire load range, PL closely tracks NL; the gap is negligible at low load and remains moderate even at high load, confirming that PL offers a good approximation at a much lower computational cost.}
\par
When comparing congestion scenarios under static pricing, the “both limited” \added{in Fig.~\ref{fig:both limited}} case performs worst: every request simultaneously stresses node and link resources, so feasible embeddings vanish faster. 
\added{Following Fig.~\ref{fig:node limited},} the “node-limited” case is the second worst (compute capacity is exhausted before bandwidth). Once candidate nodes are full, additional link capacity cannot be exploited. 
In contrast, when only links are limited, available compute still allows for some rerouting, yielding lower blocking than the node-limited case.

\subsubsection{Effect of slice type}
Fig.~\ref{fig:Blockage slice} breaks down the overall blocking probability by slice type.
We observe a clear ordering: Slice C exhibits the highest blocking rate, followed by Slice B, while Slice A is rarely blocked. This mirrors their resource/trust stringency (Table \ref{tab:slice characteristics}): Slice C requires the largest amount of resources and strictest trust constraints (tighter latency bounds, higher rate requirement, and fewer admissible operators), so feasible embeddings are exhausted first. In contrast, Slice A is lightweight and flexible, so it is typically admitted even under high load. The ranking is consistent across congestion scenarios, but the gaps widen under the “both limited” case, confirming that multi-resource scarcity amplifies the impact of demanding slices.

\subsubsection{Effect of resource price}
We analyze the effect of dynamic pricing for all congestion scenarios. The results are reported in Fig.~\ref{fig:block rate}\added{(a-d)}. Overall, introducing dynamic prices consistently lowers the blocking probability. The benefit is negligible at low load—when resources are plentiful—but becomes marked once the network approaches saturation.
This improvement is a direct consequence of adopting the Kleinrock cost function. Since this function inflates the price of a resource as its utilization $\rho$ increases (i.e., the marginal cost grows rapidly when $\rho \!\to\! 1$), incoming requests are naturally steered toward less loaded nodes/links, delaying saturation of the congested ones.
Specifically, with $\phi(\rho)=\frac{\rho}{(1-\rho)^2}$ (Kleinrock), the marginal price $\phi'(\rho)$ explodes as $\rho$ approaches 1, penalizing hotspots and incentivizing placements on lightly loaded resources.

\subsubsection{Computation time}
Table~\ref{tab:computation time} reports the average computing times (in seconds) for every request, and averaged over 50 problem instances for PL and NL at $\lambda\!\in\!\{2,4,6\}$ in the four congestion scenarios.
PL is consistently much faster: its times lie between 1.21 and 4.47\,s, whereas NL ranges from 8.50 to 28.17\,s. This corresponds to speedups of roughly $3\times$–$6\times$ (e.g., in the both-limited case at $\lambda{=}6$, 4.47\,s vs.\ 28.17\,s $\approx 6.3\times$ faster; no-congestion at $\lambda{=}6$, 2.61\,s vs.\ 17.61\,s $\approx 6.7\times$).
Moreover, NL’s runtime grows sharply with load (e.g. 8.50\,s$\rightarrow$17.61\,s in the no-congestion case), while PL remains in the low–single-digit range, showing much milder sensitivity to $\lambda$. These results confirm that PL achieves near-optimal blocking performance at a fraction of NL’s computational cost, making it practical for online operation.

\begin{table}[!t]
\centering
\caption{Average computing times (seconds) for the 50 problem instances.}
\label{tab:computation time}
\resizebox{0.6\textwidth}{!}{%
\begin{tabular}{lcccccc}
\hline
\multicolumn{1}{c}{\multirow{2}{*}{Congestion Type}} & \multicolumn{2}{c}{$\lambda=2$} & \multicolumn{2}{c}{$\lambda=4$} & \multicolumn{2}{c}{$\lambda=6$} \\ \cline{2-7} 
\multicolumn{1}{c}{}                               & PL         & NL        & PL        & NL         & PL        & NL         \\ \hline
No Congestion                                      & 1.57       & 8.5       & 1.72      & 10.54      & 2.61      & 17.61      \\
Node-limited                                       & 1.21       & 8.74      & 2.91      & 14.74      & 1.92      & 20.39      \\
Bandwidth-limited                                  & 1.49       & 9.12      & 2.32      & 12.93      & 2.76      & 18.9       \\
Both-limited                                       & 1.45       & 9.05      & 3.09      & 17.13      & 4.47      & 28.17      \\ \hline
\end{tabular}
}
\end{table}

\section{Conclusion}
\label{Conclusion}
This paper addressed online network slice deployment across multiple administrative domains under explicit trust constraints by first formulating the problem as a Node–Link (NL) integer program to obtain an optimal benchmark, and then introducing a Path–Link (PL) formulation that pre-generates trust-compliant candidate paths to drastically reduce solving time.
Simulation results over four congestion scenarios and three slice types reveal that (i) the PL formulation closely tracks the optimal NL solution while reducing computation time by roughly 3–6×, making it suitable for online operation, and (ii) dynamic resource pricing based on the Kleinrock congestion function significantly reduces blocking under scarce resources by steering traffic away from hot spots.
Future work will explore (a) learning-based path generation and pricing to further reduce runtime and tuning effort, (b) stochastic or robust formulations that explicitly model uncertainty in demand and resource availability, (c) incentive-compatible mechanisms for inter-operator trust negotiation, \added{and (d) a quantitative comparison of transitive and non-transitive trust models and their impact on feasibility, performance, and scalability.}



\ifCLASSOPTIONcaptionsoff
  \newpage
\fi

\bibliographystyle{IEEEtran}

\bibliography{biblio}

\vspace{-0.5cm}
\begin{IEEEbiography}[{\includegraphics[width=1in,height=1.25in,clip,keepaspectratio]{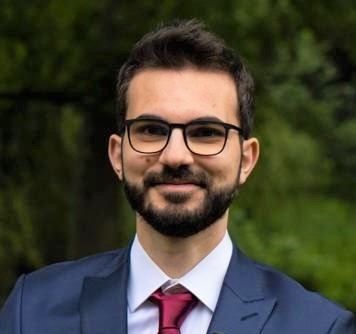}}]{Julien Ali El Amine}
received his Ph.D. degree in computer and networking from IMT Atlantique (formerly, Telecom Bretagne), Rennes, France in 2019.
He is currently an assistant professor with the College of Engineering and Technology in the American University of the Middle East (AUM), Kuwait. His research interests include green wireless communications, resource allocation, network slicing optimization, network function virtualization (NFV) and machine learning.
\end{IEEEbiography}
\vspace{-1.2cm}

\begin{IEEEbiography}[{\includegraphics[width=1in,height=1.25in,clip,keepaspectratio]{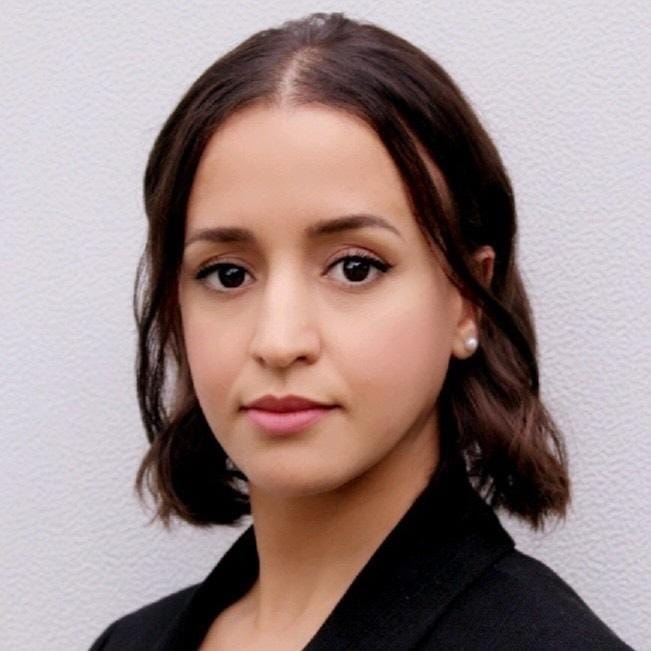}}]{Nour El Houda Nouar}
obtained her PhD in Computer Science and Telecommunications in 2022 from the University of Toulouse 1 Capitole, in collaboration with LAAS-CNRS. Her research focuses on the lifecycle management of virtualized network services (NFV), software-defined networking (SDN), and intelligent orchestration in edge-cloud environments.

\end{IEEEbiography}
\vspace{-0.0cm}

\begin{IEEEbiography}[{\includegraphics[width=1in,height=1.25in,clip,keepaspectratio]{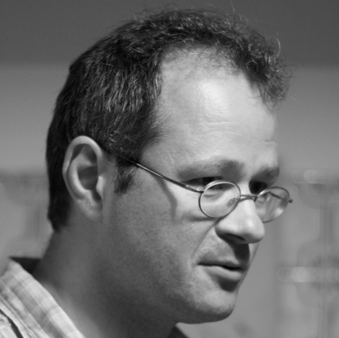}}]{Olivier Brun}
is a CNRS research staff member at LAAS, in the SARA group. He graduated from the Institut National des Télécommunication (INT, Evry, France) and he was awarded his PhD degree from Université Toulouse III (France). His research interests lie in queueing and game theories as well as network optimization.

\end{IEEEbiography}
\end{document}